# Comparing Regularized Kelvinlet Functions and the Finite Element Method for Registration of Medical Images to Sparse Organ Data

Morgan J. Ringel,[a,b] Jon S. Heiselman,[a,b,c] Winona L. Richey,[a,b] Ingrid M. Meszoely,[d] William R. Jarnagin,[c] Michael I. Miga[a,b]

[a] Vanderbilt University, Department of Biomedical Engineering, Nashville, Tennessee, USA
[b] Vanderbilt Institute for Surgery and Engineering, Nashville, Tennessee, USA
[c] Memorial Sloan-Kettering Cancer Center, Department of Surgery, New York, New York, USA
[d] Vanderbilt University Medical Center, Division of Surgical Oncology, Nashville, Tennessee, USA

Corresponding Author: Morgan J. Ringel (morgan.j.ringel@Vanderbilt.edu)
Vanderbilt University
1225 Stevenson Center Ln
Stevenson Center 5824
Nashville, TN 37240

# Highlights

- Regularized Kelvinlets are used to model linear elasticity in soft tissue organs
- Images are registered to sparse geometric data using regularized Kelvinlets or FEM
- Registration accuracy is comparable between regularized Kelvinlets and FEM
- Regularized Kelvinlets registration reduces computation time compared to FEM

# Abstract

Image-guided surgery collocates patient-specific data with the physical environment to facilitate surgical decision making in real-time. Unfortunately, these guidance systems commonly become compromised by intraoperative soft-tissue deformations. Nonrigid image-to-physical registration methods have been proposed to compensate for these deformations, but intraoperative clinical utility requires compatibility of these techniques with data sparsity and temporal constraints in the operating room. While linear elastic finite element models are effective in sparse data scenarios, the computation time for finite element simulation remains a limitation to widespread deployment. This paper proposes a registration algorithm that uses regularized Kelvinlets, which are analytical solutions to linear elasticity in an infinite domain, to overcome these barriers. This algorithm is demonstrated and compared to finite element-based registration on two datasets: a phantom dataset representing liver deformations and an *in vivo* dataset representing breast deformations. The regularized Kelvinlets algorithm resulted in a significant reduction in computation time compared to the finite element method. Accuracy as evaluated by target registration error was comparable between both methods. Average target registration errors were $4.6 \pm 1.0$ and $3.2 \pm 0.8$ mm on the liver dataset and $5.4 \pm 1.4$ and $6.4 \pm 1.5$ mm on the breast dataset for the regularized Kelvinlets and finite element method models, respectively. This work demonstrates the generalizability of using a regularized Kelvinlets registration algorithm on multiple soft tissue elastic organs. This method may improve and accelerate registration for image-guided surgery applications, and it shows the potential of using regularized Kelvinlets solutions on medical imaging data.

# 1 Introduction

Medical image-to-physical registration is the process of solving for a transformation to align imaging space to the physical patient space and establish anatomical correspondence (Wang and Li, 2019). When medical imaging data is properly aligned to the physical orientation of the patient at the time of surgery, surgeons can integrate that information to adjust surgical decision making in real-time (Alam et al., 2018). While potentially powerful, image-to-physical registration for image-guided surgery (IGS) is often accomplished using sparse geometric data that can only be acquired in the surgical environment at the time of surgery. This technology has become standard of care for neurosurgery with typical IGS systems featuring either an optical or electromagnetic tracking system, a visual display, and a computer to handle the processes associated with data acquisition, processing, and mathematical transformations (Sorriento et al., 2020). Research into the translation of IGS into other soft-tissue organs is ongoing. For example, image guidance is also of interest for hepatic surgeries in order to successfully navigate liver vascular structures, localize tumors, and safely perform resection (Gavriilidis et al., 2022). Although more speculative, image guidance for lumpectomy surgeries in the breast is also being explored to inform total tumor excision and reduce reoperation rates (Barth et al., 2019; Conley et al., 2015; Joukainen et al., 2021).

While rigid registration to the physical environment can be informative, the accuracy of using preoperative imaging within clinical workflows is diminished by nonrigid soft tissue deformations that occur during surgery (Payan, 2012). Using nonrigid registration to compensate for this deformation during IGS is challenging because of the limited amount of intraoperative data, which conventionally consists of sparse point clouds or landmarks localized on the organ surface. Furthermore, nonrigid registration algorithms must be near real-time so as to not significantly prolong surgical time. Thus, nonrigid image registration approaches that are compatible with sparse data constraints and the intraoperative workflow are essential to the complete realization of IGS registration methods. Biomechanical models can be used to constrain the transformation vector field to comply with the physical constraints associated with soft tissue deformation. A well-established biomechanical modeling approach for estimating soft tissue deformations is the use of Hookean elasticity theory and the finite element method (FEM). FEM involves discretizing the organ volume into a multitude of subdomains, i.e., finite elements, where weak forms of the governing partial differential equations are integrated and assembled into large systems of coupled equations (Bower, 2009). However, FEM modeling can be cumbersome as it requires a large-scale matrix assembly, matrix inversion, and user-prescribed known boundary condition designation to solve for a displacement field.

Recent work has focused on developing biomechanical modeling approaches to enable soft tissue dynamic simulations in real-time for use in surgical training and guidance applications. For example, the open-source SOFA package was developed as a modular framework for computationally fast medical simulations (Faure et al., 2012). GPU implementations of SOFA have enabled real-time surgical simulations for multiple applications like endovascular intervention procedures and cataract surgery (Comas et al., 2008; Guo et al., 2018). Another GPU-enabled open-source package called NiftySim uses the total Lagrangian explicit dynamics algorithm to achieve real-time biomechanical simulations and has been demonstrated in the breast and prostate (Johnsen et al., 2015). While both the SOFA and NiftySim packages enable real-time simulations, they still require involved and prescribed boundary condition designation. Also, since these packages have mostly been used in simulation, in-depth analyses of registration accuracy for IGS using these methods are limited. Machine learning methods that use FEM simulations as training data have also been explored for real-time biomechanical modeling (Phellan et al., 2020). However, these methods still require computational time and resources for training, and generalizability of these methods to novel organ geometries or loading conditions can be a concern depending on the training data (Mendizabal et al., 2020; Pellicer-Valero et al., 2020).

This work builds upon two previously published biomechanical modeling methods. The first method is the regularized Kelvinlets solutions proposed by Goes *et al*. in 2017 (de Goes and James, 2017). The regularized Kelvinlets solutions are described as sculpting brushes used by animation artists in a forward-solve manner where displacements are computed given input forcing vectors. They are computationally less involved than other biomechanical models while still being derived from the fundamental three-dimensional (3D) equations for linear elasticity. The second method is the linearized iterative boundary reconstruction (LIBR) method proposed by Heiselman *et al*. in 2020 (Heiselman et al., 2020). The LIBR method uses an inverse-problem approach to reconstruct a registered deformation state from a linear basis of displacement modes generated with an FEM displacement model. The work here proposes a synthesis of these methods using analytic regularized Kelvinlets displacement solutions for sparse data organ registration. Regularized Kelvinlets are leveraged in an inverse-problem registration scheme where forcing vectors are reconstructed such that the total deformation response matches a set of given sparse data displacement inputs. To investigate the rigor of the approach, the method is applied to two datasets to show generalizability to multiple elastic organ models: an elastic phantom dataset representing approximately 112 sparse data liver deformations, and an *in vivo* breast dataset representing breast deformations from 7 volunteers. To demonstrate that regularized Kelvinlets offer a competitive alternative for elastic registration problems with limited data, registration results are compared to the FEM-based counterpart presented in Heiselman *et al.* (Heiselman et al., 2020). Computation time, target accuracy, and image deformation results are evaluated and compared for the two registration methods and datasets.

## 2  Methods
### *2.1  Algorithm Overview*
An overview of the sparse data registration method is shown in Figure 1. The sparse data registration method features two phases: a precomputation phase (Figure 1, upper box) where a basis of displacement solutions is generated on a specific organ geometry, and a reconstruction phase (Figure 1, lower box) where the displacement basis is used to deform the organ to match sparse data inputs. First, the regularized Kelvinlets displacement solutions (Figure 1, purple) detailed in (de Goes and James, 2017) are proposed for generating a realistic biomechanical displacement basis in the precomputation phase. To assess fidelity, the regularized Kelvinlets displacement basis is compared to a displacement basis generated using a more conventional FEM model (Figure 1, green) from previous work in (Heiselman et al., 2020). Finally, the algorithm for the reconstruction phase from (Heiselman et al., 2020) is used for both biomechanical model displacement bases realizations for direct comparison (Figure 1, lower box). Equation notation is written such that constants are italicized, vectors are bolded, and matrices are double-struck letters.

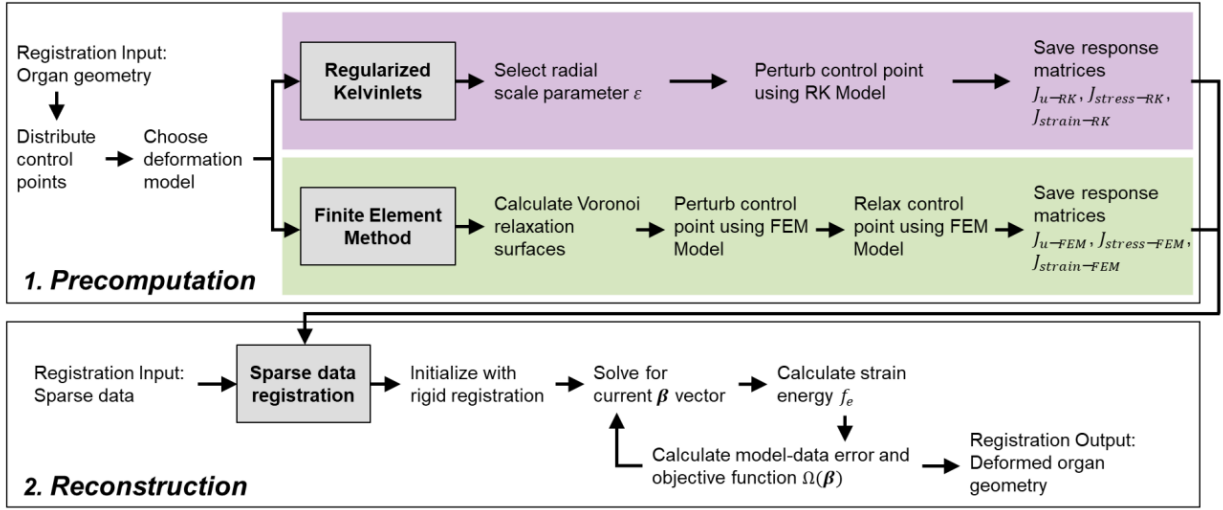

*Figure 1 – A flowchart overview of the sparse data registration method. The method is divided into a precomputation phase (1, upper box) and a reconstruction phase (2, lower box). Either the Regularized Kelvinlets (purple) or the Finite Element Method (green) can be used as the biomechanical model to generate the displacement basis used for reconstruction.*

### 2.1.1 Regularized Kelvinlets Displacement Solution

The regularized Kelvinlets displacement solution is derived from the Navier-Cauchy equations, which govern the relationship between displacement and applied force at static equilibrium for an isotropic, linear elastic material.

$$\frac{E}{2(1+v)}\nabla^2 \mathbf{u} + \frac{E}{2(1+v)(1-2v)}\nabla(\nabla \cdot \mathbf{u}) + \mathbf{F} = 0 \tag{1}$$

The Navier-Cauchy equations are shown in (1) where $E$ is Young's modulus, $v$ is Poisson's ratio, $\mathbf{u}$ is the displacement vector, and $\mathbf{F}$ is the applied force. A Kelvinlet (also known as Kelvin's state or Kelvin solution), is a fundamental solution of linear elasticity corresponding to a singular point load $\delta(\mathbf{x} - \mathbf{x_0})$ applied to an infinite elastic space. The displacement field $\mathbf{u}(x)$ can be determined by finding the Green's function for the Navier-Cauchy equations resulting in the Kelvinlet equation shown below.

$$\mathbf{u}(\mathbf{r}) = \left[\frac{a-b}{r}\mathbb{I} + \frac{b}{r^3}\mathbf{r}\mathbf{r}^t\right]\mathbf{f} \tag{2}$$

In (2), $\mathbf{r} = \mathbf{x} - \mathbf{x_0}$ is the position vector from the load location $\mathbf{x_0}$ to the observing location $\mathbf{x}$ with $r = \|\mathbf{r}\|$. The coefficients are equal to $a = \frac{(1+v)}{2\pi E}$ and $b = \frac{a}{[4(1-v)]}$.

In the Kelvinlet equation (2), the displacement solution $\mathbf{u}(\mathbf{r})$ becomes indefinite as $r$ approaches zero. To address this, instead of solving for the displacement solution to a singular point load $\delta(\mathbf{x} - \mathbf{x_0})$, a "bump" or "cutoff" function is used for regularization. The smoothed forcing function was first proposed in (Cortez et al., 2005) for regularized Stokeslets in fluid simulation and can be represented as $\mathbf{f}\rho_\varepsilon(\mathbf{r})$ where $\rho_\varepsilon(\mathbf{r})$ is the normalized density function shown in (3).

$$\rho_\varepsilon(\mathbf{r}) = \frac{15\varepsilon^4}{8\pi}\frac{1}{r_\varepsilon^7} \tag{3}$$

In (3), $r_\varepsilon = \sqrt{r^2 + \varepsilon^2}$ is the regularized distance, and $\varepsilon$ is the radial scale of regularization. Solving for the analytical solution with the above smoothed forcing function yields the formula for regularized Kelvinlets proposed in (de Goes and James, 2017) and reproduced below in (4).

$$\mathbf{u}_\varepsilon(\mathbf{r}) = \left[\frac{a-b}{r_\varepsilon}\mathbb{I} + \frac{b}{r_\varepsilon^3}\mathbf{r}\mathbf{r}^t + \frac{a}{2}\frac{\varepsilon^2}{r_\varepsilon^3}\mathbb{I}\right]\mathbf{f} \tag{4}$$

The solution in (4) assumes an infinite elastic medium, and it does not account for the geometry of the domain.

The *r* vectors and corresponding **u**$_\varepsilon$(*r*) vectors from a regularized Kelvinlet displacement solution can be computed for every individual node in an organ mesh. The **u**$_\varepsilon$(*r*) vectors can be concatenated to form a displacement vector ***d***$_{RK}$ of length 3*M* where *M* is the number of nodes in the mesh shown in (5).

$$d_{RK} = \begin{bmatrix} u_\varepsilon(r_1) \\ u_\varepsilon(r_2) \\ \ldots \\ u_\varepsilon(r_M) \end{bmatrix} \quad (5)$$

*2.1.2   Finite Element Method Displacement Solution*

The regularized Kelvinlet displacements are compared to displacement solutions derived from a conventional elastic FEM model. For FEM model realization, organ volumes are discretized with tetrahedral meshes with 4 mm element edge lengths. The Galerkin weighted residual method with linear Lagrange basis functions is applied to integrate the partial differential equations associated with the Navier Cauchy equations in (1). This process produces a coupled set of differential equations as shown in (6).

$$\mathbf{u_{FEM}} = \mathbb{K}^{-1}\mathbf{F} \quad (6)$$

In (6), $\mathbb{K}$ is the global stiffness matrix, $\mathbf{u_{FEM}}$ is the vector of displacements, and **F** is the vector containing forcing terms and boundary conditions.

To create a non-singular FEM displacement solution response to a distributed point load, the methodology described in (Heiselman et al., 2020) perturbs and relaxes a singular boundary mesh point, rendering a comparable solution to regularized Kelvinlets. For this displacement solution, a series of control points are distributed evenly on the control surface of the organ. The control point at the load location *x$_0$* is perturbed while all other control points are fixed. After perturbation, the Voronoi tile region surrounding the point load location is relaxed, and a distributed load that results in identical far-field displacements is computed in accordance with the Saint-Venant principle. The resulting displacement vector after relaxation is denoted as ***d***$_{FEM}$ which is of length 3*M* where *M* is the number of nodes in the mesh.

The displacement and strain norm fields from the relaxed FEM displacement solution and the regularized Kelvinlets displacement solution at one point load location on liver and breast organs are shown in Figure 2. Two values of the radial scale parameter *ε* are shown for the regularized Kelvinlets displacement solutions. Both the relaxed FEM and regularized Kelvinlets displacement solutions are approximations of deformation responses to point load forces. As shown in Figure 2, both solutions exhibit a local response centered on the point at which the perturbation is applied. However, the relaxed FEM displacement solution accounts for organ geometry, and it is also influenced by all other control point locations that are fixed during perturbation. The regularized Kelvinlet solution is computed as if the organ is embedded in an infinite elastic medium, and it does not account for organ specific geometry.

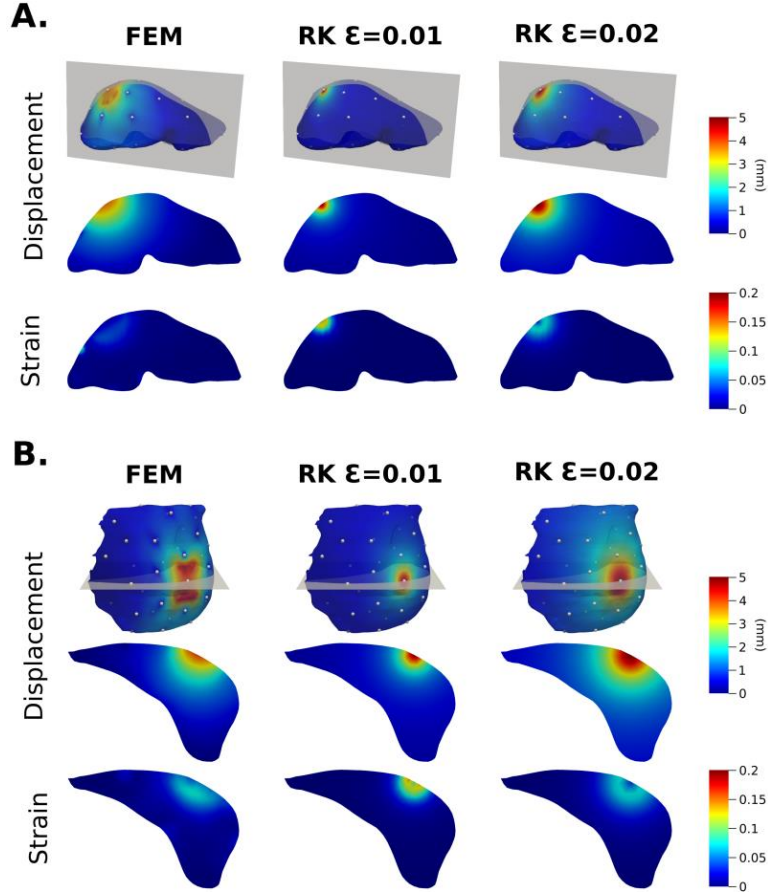

*Figure 2 – Comparison of displacement and strain norm fields generated from perturbation of a singular control point using the FEM method and the Regularized Kelvinlets (RK) method with two values of the radial scale parameter ε. Fields are shown on liver geometry (A) and breast geometry (B). Each field is normalized to have a maximum displacement value of 5 mm.*

### 2.1.3 Sparse Data Registration

From sections 2.1.1 and 2.1.2 above, two linear elastic modeling methods have been presented for representing the displacements caused by a distributed point load force. The linearized iterative boundary reconstruction (LIBR) method, detailed in (Heiselman et al., 2020), uses a displacement basis composed of FEM displacement solutions and solves for an optimal linear combination of these solution vectors to minimize model-data error and achieve a sparse-data-driven nonrigid registration. In this work, the method proposed in (Heiselman et al., 2020) (referred to as LIBR+FEM) has been re-engineered to employ regularized Kelvinlets displacement solutions as the displacement basis (referred to as LIBR+RK) in lieu of FEM. Both LIBR+FEM and LIBR+RK displacement bases methods are linearly superposed and combined within an optimization framework to recover an optimal registration displacement state.

To create the LIBR+FEM displacement basis, each control point is perturbed and relaxed in the *x, y,* and *z* directions to create a series of $d_{FEM}$ basis vectors. In total, $3k$ basis vectors are generated where $k$ is the number of control points. As detailed in (Heiselman et al., 2020), the registered deformation state $\tilde{u}_{FEM}$ can be approximated using (7).

$$\tilde{u}_{FEM} = \mathbb{J}_{u-FEM}\, \alpha_{FEM} = \begin{bmatrix} | & | & & | \\ d_{FEM-1} & d_{FEM-2} & \ldots & d_{FEM-3k} \\ | & | & & | \end{bmatrix} \alpha_{FEM} \qquad (7)$$

In (7), $\tilde{u}_{FEM}$ approximates the deformation state based on the linear combination of the displacement basis functions written as the displacement response matrix $\mathbb{J}_{u-FEM}$ (size $3M \times 3k$), and $\alpha_{FEM}$ is the displacement basis function weights of length $3k$. Along with displacement, a $6M \times 3k$ stress response matrix $\mathbb{J}_{stress-FEM}$ and $6M \times 3k$ strain response matrix $\mathbb{J}_{strain-FEM}$ are also formulated.

To create an analogous LIBR+RK displacement basis, the forcing vectors $\boldsymbol{f} = \begin{bmatrix}1\\0\\0\end{bmatrix}$, $\boldsymbol{f} = \begin{bmatrix}0\\1\\0\end{bmatrix}$, and $\boldsymbol{f} = \begin{bmatrix}0\\0\\1\end{bmatrix}$ are used in (4) to perturb each control point in the *x, y,* and *z* directions using the regularized Kelvinlets displacement solutions. This results in 3k $\boldsymbol{d_{RK}}$ basis vectors for formulating (8).

$$\tilde{\mathbf{u}}_{\mathbf{RK}} = \mathbb{J}_{u-RK}\,\boldsymbol{\alpha_{RK}} = \begin{bmatrix} | & | & & | \\ d_{RK-1} & d_{RK-2} & \cdots & d_{RK-3k} \\ | & | & & | \end{bmatrix} \boldsymbol{\alpha_{RK}} \tag{8}$$

Like in (7), $\tilde{\mathbf{u}}_{\mathbf{RK}}$ approximates the deformation state using the regularized Kelvinlets model with $\mathbb{J}_{u-RK}$, $\mathbb{J}_{stress-RK}$, and $\mathbb{J}_{strain-RK}$ being the displacement, stress, and strain response matrices and $\boldsymbol{\alpha_{RK}}$ being the weight vector.

The registration task is to solve for the optimal $\boldsymbol{\alpha_{FEM}}$ or $\boldsymbol{\alpha_{RK}}$ vector, combined with rigid transformation parameters $\boldsymbol{\tau}$ (translation) and $\boldsymbol{\theta}$ (rotation), that minimizes model-data error and the strain energy of the deformation. The approach proposed in (Heiselman et al., 2020) for formulating the objective function, establishing correspondences for calculating model-data error, incorporating a rigid transformation, and calculating the strain energy term is kept identical for the LIBR+FEM and LIBR+RK methods. This algorithmic consistency allows for a direct comparison of the two biomechanical models without the influence of other variables. The objective function as described in (Heiselman et al., 2020) is shown in (9).

$$\Omega(\boldsymbol{\beta}) = \sum_F \frac{w_F}{N_F} \sum_{i=1}^{N_F} f_i^2 + w_E f_E^2 \tag{9}$$

In (9), $\boldsymbol{\beta}=[\boldsymbol{\alpha_{FEM}}, \boldsymbol{\tau}, \boldsymbol{\theta}]$ or $\boldsymbol{\beta}=[\boldsymbol{\alpha_{RK}}, \boldsymbol{\tau}, \boldsymbol{\theta}]$ depending on the chosen basis functions. $f_i$ represents the model-data error at datapoint $i$, $w_F$ represents the weight of data feature $F$, and $N_F$ represents the total number of data points in feature $F$. A strain energy regularization term is included in $\Omega(\boldsymbol{\beta})$ with $f_E$ being the average strain energy of a deformation state and $w_E$ being the strain energy regularization weight. $f_E$ is computed with either (10) or (11) depending on the method (LIBR+FEM or LIBR+RK).

$$f_E = \frac{1}{2M}\boldsymbol{\alpha_{FEM}}^{\mathrm{T}}\left(\mathbb{J}_{strain-FEM}^{\mathrm{T}}\mathbb{J}_{stress-FEM}\right)\boldsymbol{\alpha_{FEM}} \tag{10}$$

$$f_E = \frac{1}{2M}\boldsymbol{\alpha_{RK}}^{\mathrm{T}}\left(\mathbb{J}_{strain-RK}^{\mathrm{T}}\mathbb{J}_{stress-RK}\right)\boldsymbol{\alpha_{RK}} \tag{11}$$

The data features are chosen to mimic data that could theoretically be acquired in a surgical environment for intraoperative registration and are discussed in more detail for each individual organ registration in section 2.2. Levenberg-Marquardt optimization is used to iteratively solve for $\boldsymbol{\beta}$ with a termination criterion of $|\Delta\Omega(\boldsymbol{\beta})| < 10^{-12}$.

### 2.2 *Datasets and Experimentation*

Two independent datasets were used to evaluate and compare the LIBR+FEM and LIBR+RK methods. The first is a deformation dataset generated from an elastic silicone liver phantom. The second is a deformation dataset generated from supine breast magnetic resonance (MR) imaging from healthy female volunteers.

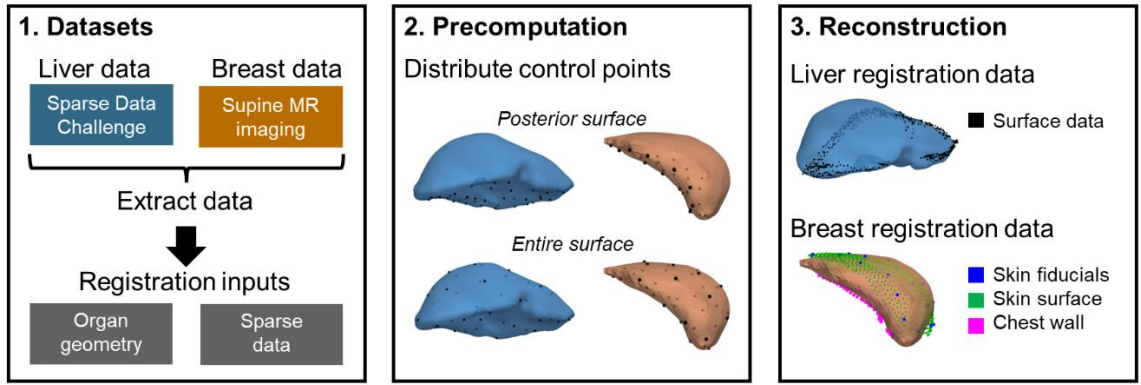

*Figure 3 – Data processing steps necessary for registration using the liver (blue) and breast (orange) datasets. Organ geometry and sparse data are used as registration inputs (left). Control points are placed on either the posterior surface or entire surface of the organ mesh during the precomputation phase (middle). Sparse surface data is used for registration during the reconstruction phase (right).*

*2.2.1   Phantom Liver Dataset*

The phantom liver dataset was obtained from the Sparse Data Challenge for image-guided liver surgery (Brewer et al., 2019; Collins et al., 2017). The challenge dataset includes a liver geometry mesh (Figure 3, middle, blue), 112 sparse data feature configurations (Figure 3, right, blue), and 159 targets. The sparse data feature configurations were generated from contact and non-contact intraoperative data collections. The liver phantom was composed of a silicone material cast from a mold that was obtained from a human patient CT image volume and emulated the stiffness of liver tissue. Deformations were applied to the phantom's posterior surface. Parameter values of $E = 2100$ Pa, $v = 0.45$, and $w_E = 10^{-8}$ Pa$^{-2}$ were used for both the LIBR+FEM and LIBR+RK methods based on parameter values used in previous work (Heiselman et al., 2020).

*2.2.2   In Vivo Breast Dataset*

The *in vivo* breast dataset was acquired from supine MR breast imaging of healthy volunteers used in previous work (Richey et al., 2022; Ringel et al., 2022). 26 MR-visible skin surface fiducials were placed on the breast prior to imaging. Volunteers were instructed to place their arm by their side (arm-down configuration) followed by raising their arm above their head (arm-up configuration), which caused deformation of the ipsilateral breast. This deformation was meant to reproduce the types of breast deformations that occur due to positional changes during lumpectomy surgeries. MR imaging was acquired in the arm-down and arm-up configurations. Eleven breast imaging sets were included in the dataset. Breasts in the arm-down configuration images were segmented and used to create 11 unique tetrahedral meshes (Figure 3, middle, orange). Approximately 20 corresponding subsurface targets were identified in each image pair using the glandular features in the breast. The sparse data feature configurations included the MR visible skin fiducial points, a sparse skin surface point cloud, and a sparse chest wall surface point cloud (Figure 3, right, orange). These features simulated data that could be acquired using a previously proposed breast navigation image-guidance research system (Richey et al., 2021). Methodology for using these sparse data sources with the objective function in (9) for registration is discussed in (Richey et al., 2022). Parameter values of $E = 2100$ Pa, $v = 0.45$, and $w_E = 10^{-9}$ Pa$^{-2}$ were used for both the LIBR+FEM and LIBR+RK methods based on parameters used in previous work (Richey et al., 2022). A comparison summary of both datasets (liver and breast) is reported in Table 1.

*Table 1 – Summary table of the datasets used to evaluate and compare registration methods.*

|  | **Precomputation Data** | | **Reconstruction Data** | | | |
|---|---|---|---|---|---|---|
|  | # Geometries | Mesh size | Feature data types | # Feature configurations | # Feature data points | # Evaluation Targets |
| Liver Dataset | 1 | 29,545 nodes 159,014 elements | 1. Surface data | 112 | 1,151 ± 308 | 159 |
| Breast Dataset | 11 | 10,174 ± 3,086 nodes 50,753 ± 17,026 elements | 1. Fiducial points 2. Skin surface Data 3. Chest wall surface data | 11 (1 per geometry) | 1,243 ± 910 | 22 ± 3 (237 total) |

*2.2.3 Experimental Setup*

Registrations using the LIBR+FEM and LIBR+RK methods were performed on both organ datasets (liver and breast). For both datasets, two control point distribution strategies were tested: placing control points distributed evenly just on the posterior surface of the organ and placing control points distributed evenly everywhere on the organ surface (Figure 3, middle). For the breast dataset, the posterior surface refers to the intersecting surface between the chest wall and breast parenchyma. K-means clustering was used to distribute control points evenly, and a parameter sweep of the number of control points $k$ was performed iterating between 10-190 in increments of 30. For the LIBR+RK method, a parameter sweep of the radial scale parameter $\varepsilon$ was performed iterating through the values $\varepsilon$ = [0.001, 0.002, 0.005, 0.01, 0.02, 0.05, 0.1, 0.2, 0.5] meters.

Registration performance was evaluated by calculating the root mean squared (RMS) target registration error (TRE), which is the root mean squared error of all individual target errors in one registration case. Computation times for each method were also compared. The precomputation phase for the LIBR+FEM method was parallelized on 8 threads of an AMD Ryzen 7 3700X CPU. All other computations (LIBR+FEM reconstruction and LIBR+RK precomputation and reconstruction) were performed on a single thread of an AMD Ryzen 7 3700X CPU.

## 3 Results

### 3.1 Computation Time

The average computation times for registrations performed using the LIBR+FEM (green) and the LIBR+RK (purple) methods on the breast and liver datasets are shown in Figure 4. Times are reported for the precomputation phase (left), reconstruction phase (middle), and total registration time (right), which is the sum of the time for both phases. Control points were placed on the posterior surface of the organ for the times reported for both methods, and a radial scale parameter value $\varepsilon = 0.01$ was used for the LIBR+RK registrations. For the LIBR+FEM method, the precomputation phase accounts for most of the total computation time. For both methods, total computation time scales approximately linearly with the number of control points. Because the precomputation phase takes significantly less time with the LIBR+RK method than with the LIBR+FEM method, the total computation time is reduced by an average of 4.2–34.4 minutes for the liver dataset and an average of 1.4–10.4 minutes for the breast dataset depending on the number of control points.

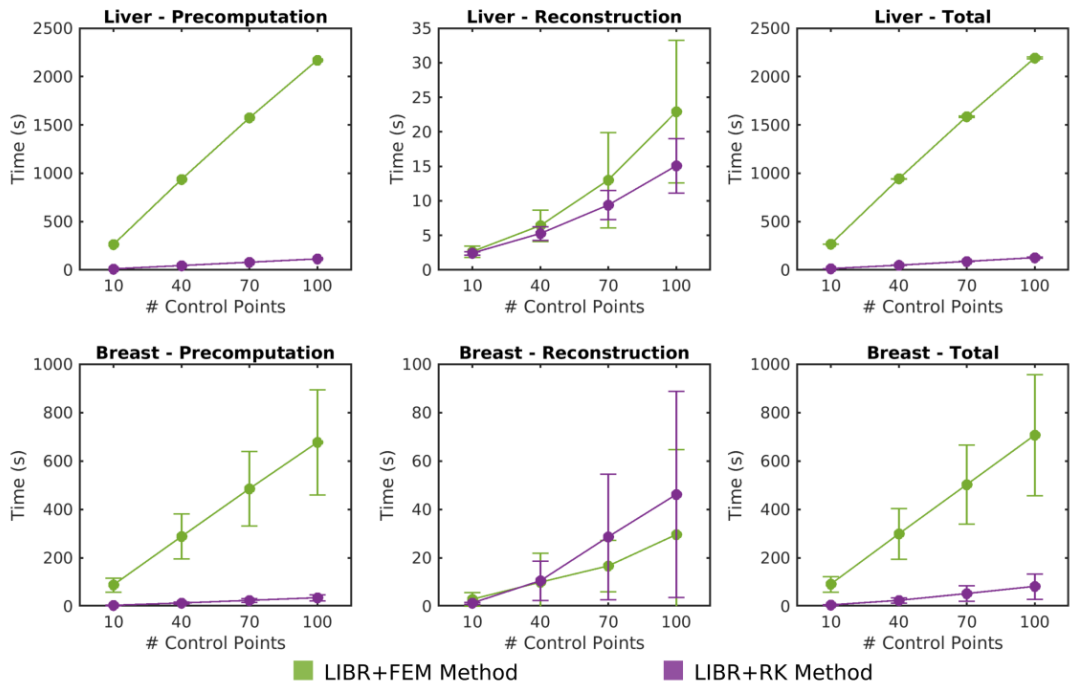

*Figure 4 –Computation time for registration using the LIBR+FEM (green) and LIBR+RK (purple) methods. Error bars represent the standard deviation, and no error bars are reported for the liver precomputation phase because there was only one geometry in the dataset. Control points were placed on the posterior organ surface for both methods, and ε=0.01 for the LIBR+RK method.*

### 3.2 Parameter Sweep Results

The registration accuracy results from the parameter sweep are shown in Figure 5. Each square denotes the average RMS TRE for a parameter sweep combination averaged across either 112 data configurations for the liver dataset or across 11 volunteer geometries for the breast dataset. Average RMS TRE is displayed as a function of registration method (LIBR+FEM or LIBR+RK), control point placement strategy (either distributed on the organ's posterior surface or on the entire organ surface), radial scale $\varepsilon$ for LIBR+RK registrations (x-axis), and number of control points (y-axis).

The optimal number of control points and radial scale $\varepsilon$ parameters for each method, control point placement, and dataset are reported in Table 2. The overall optimal registration for the liver dataset was with the LIBR+FEM method with $k = 40$ control points distributed on the posterior surface, resulting in an average RMS TRE of 3.2 ± 0.8 mm. The best registration accuracy for the liver dataset with the LIBR+RK method was 4.6 ± 1.0 mm which occurred with $k = 160$ control points distributed everywhere with $\varepsilon = 0.01$. For the breast dataset, the optimal registration was with the LIBR+RK method with $k = 70$ control points and $\varepsilon = 0.05$ with control points distributed on the posterior surface, resulting in an average RMS TRE of 5.4 ± 1.4 mm. With the LIBR+FEM method, the optimal registration was 6.4 ± 1.5 mm with $k = 40$ control points distributed on the posterior surface.

Examining the parameter sweep overall, many of the parameter combinations resulted in average RMS TRE values that were within a 1 mm range of the optimal RMS TRE values. These parameter combinations are indicated in Figure 5 by the inner black borders. Broadly speaking, for the number of control points, the optimal parameter combinations for the LIBR+RK implementations typically had a higher number of control points than the LIBR+FEM implementations. For the radial scale parameter $\varepsilon$, the extremes of the parameter sweep resulted in worse average RMS TRE showing the importance of selecting an $\varepsilon$ parameter appropriate for the scale of the geometric organ. When comparing the LIBR+FEM and LIBR+RK methods for both datasets, average registration accuracy generally fell within the 4-7 mm range for most parameter combinations, with the exception being the registration accuracy on the liver dataset using the LIBR+FEM method which resulted in accuracies in the 3-4 mm range.

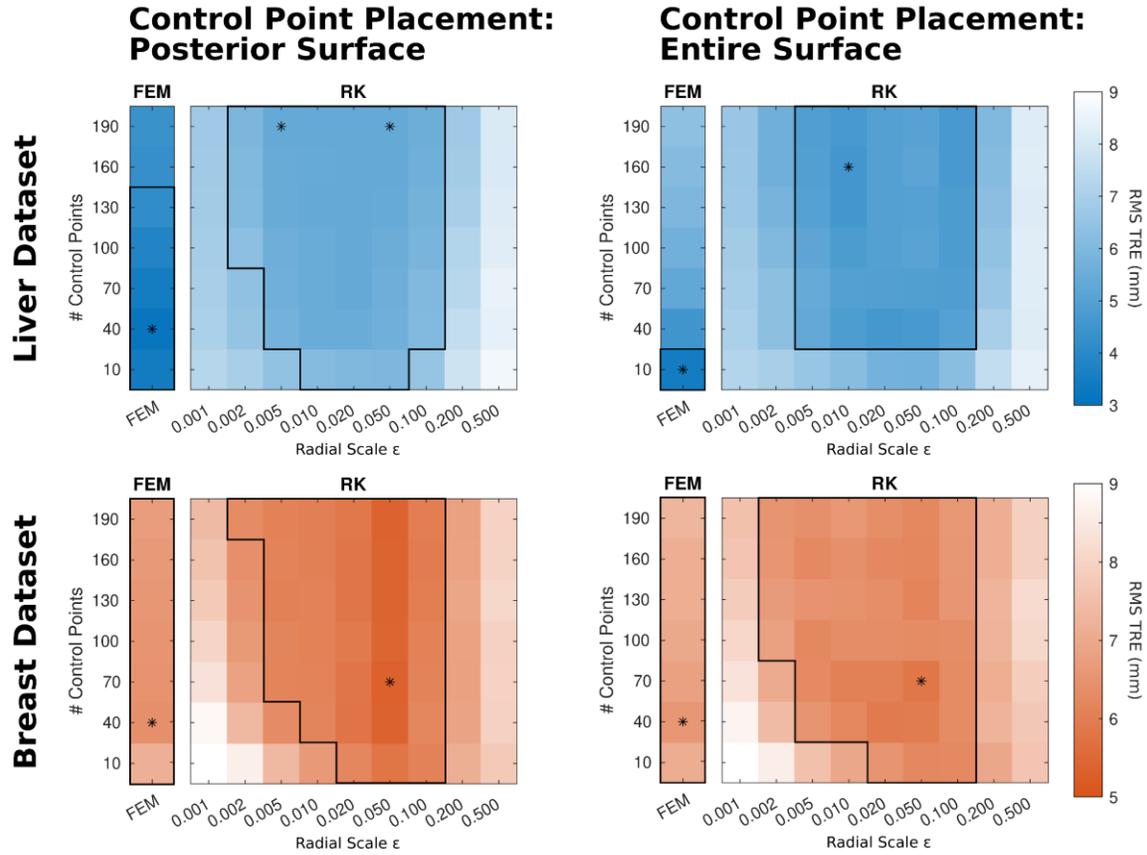

*Figure 5 – Parameter sweep results for the liver (blue) and breast (orange) datasets. Control points were distributed on the posterior (left) and entire (right) surface of the organ. Average RMS TRE values are plotted as a function of the number of control points and radial scale parameter ε. Optimal average RMS TRE values are denoted by asterisks, and numerical values are reported in Table 2. The black border denotes areas where RMS TRE values are within 1 mm of the minimum.*

*Table 2 – Table showing the optimal average RMS TRE values resulting from the parameter sweep.*

|  | Control Point Placement | Method LIBR+{FEM or RK} | Optimal Average RMS TRE ± std (mm) | # Control Points | Radial Scale ε |
|---|---|---|---|---|---|
| Liver Dataset | Posterior surface | FEM | 3.2 ± 0.8 | 40 | - |
|  |  | RK | 5.4 ± 1.1 | 190 | 0.005, 0.05 |
|  | Entire surface | FEM | 3.5 ± 0.9 | 10 | - |
|  |  | RK | 4.6 ± 1.0 | 160 | 0.01 |
| Breast Dataset | Posterior surface | FEM | 6.4 ± 1.5 | 40 | - |
|  |  | RK | 5.4 ± 1.4 | 70 | 0.05 |
|  | Entire surface | FEM | 6.6 ± 1.9 | 40 | - |
|  |  | RK | 5.9 ± 1.7 | 70 | 0.05 |

### 3.3 Registration Accuracy

The optimal parameters for the LIBR+FEM and LIBR+RK registration methods for each dataset were then used to compare organ deformations and individual target errors for example cases. The following four optimal registrations were examined: 1) $k = 40$, posterior surface distribution for LIBR+FEM method on the liver dataset, 2) $k = 160$, $\varepsilon = 0.01$, entire surface distribution for the LIBR+RK method on the liver dataset, 3) $k = 40$, posterior surface distribution for the LIBR+FEM method on the breast dataset, and 4) $k = 70$, $\varepsilon = 0.05$, posterior surface distribution for the LIBR+RK method on the breast dataset. For the liver dataset, individual target ground-truth locations for 35 targets in 4 out of the 112 deformations (deformations 44, 57, 67, and 84 denoted as L1-L4) are provided as a part of the Sparse Data Challenge for visualization, calculation, and examination of individual target errors. For the breast dataset, a small, medium, and large breast volume examples (denoted as B1-B3 respectively) are selected for examining organ deformation and individual target error.

The global organ deformations for the LIBR+FEM (green) and the LIBR+RK (purple) methods are shown on the liver (A) and breast (B) datasets in Figure 6. The registered deformation states using both methods appear qualitatively similar. The largest apparent discrepancy in global deformation is shown in the largest volume breast case B3. In addition to using RMS TRE as a metric for evaluating registration performance, comparing the organ mesh shape after registration confirms that both methods are producing similar global deformations.

Target errors were also examined on an individual dataset level in Figure 7. For the 4 example liver data configurations (L1-L4), the LIBR+FEM registrations performed slightly better than the LIBR+RK registrations. For the 3 example breast datasets (B1-B3), the LIBR+RK registrations performed slightly better than the LIBR+FEM registrations. Overall, most individual target errors were below 10 mm with several maximum target errors exceeding 10 mm.

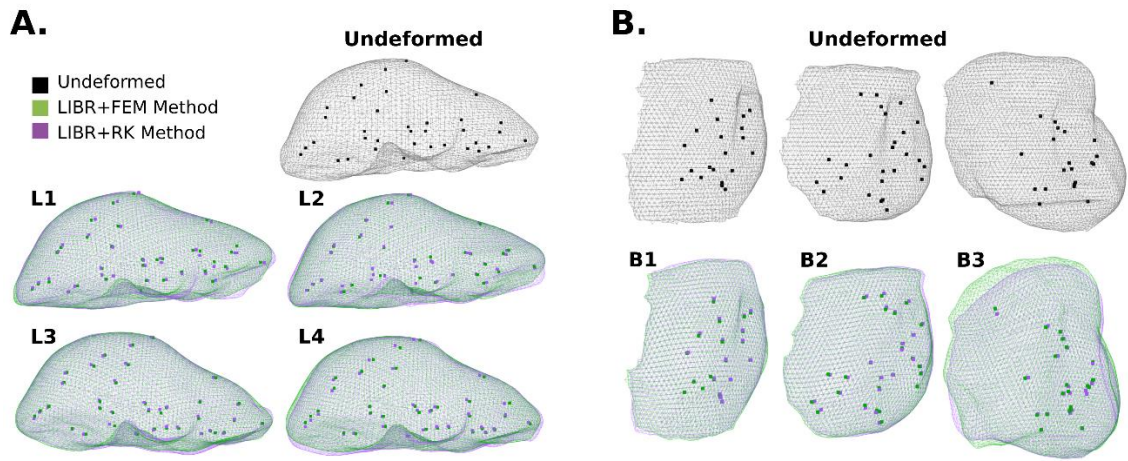

*Figure 6 – Registered organs and targets from (A) 4 example liver registrations (L1-L4) and (B) 3 example breast registrations (B1-B3). The undeformed geometry (black) is compared to the deformed geometry using the LIBR+FEM method (green) and the LIBR+RK method (purple).*

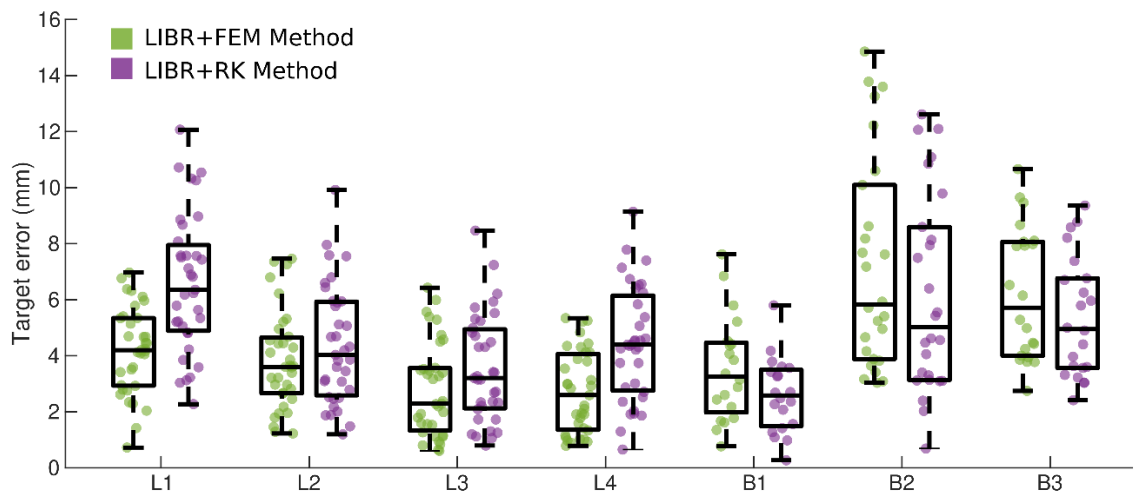

*Figure 7 – Individual target errors from 4 example liver registrations (L1-L4) and 3 example breast registrations (B1-B3) comparing the LIBR+FEM method (green) and the LIBR+RK method (purple). Boxplot whiskers indicate minimum and maximum target errors.*

## 3.4 Image Deformation

For image-guided surgery, deforming a volumetric image may be useful for visualizing the deformed anatomy. As noted above, the Sparse Data Challenge liver phantom geometry was derived from a patient's liver CT image volume. Despite deformations being based on a silicone phantom counterpart, the original liver image volume can be deformed using the deformation fields from our applied registration methods as a means to compare image similarity among all imaging volumes (original and deformed). This process is shown in Figure 8 for the 4 example liver data configurations for the LIBR+FEM method (green), the LIBR+RK method (purple), and the undeformed original organ for reference (yellow). Vasculature features within the liver segmentations for the deformed images look qualitatively similar when comparing the resulting deformed images from the two registration methods.

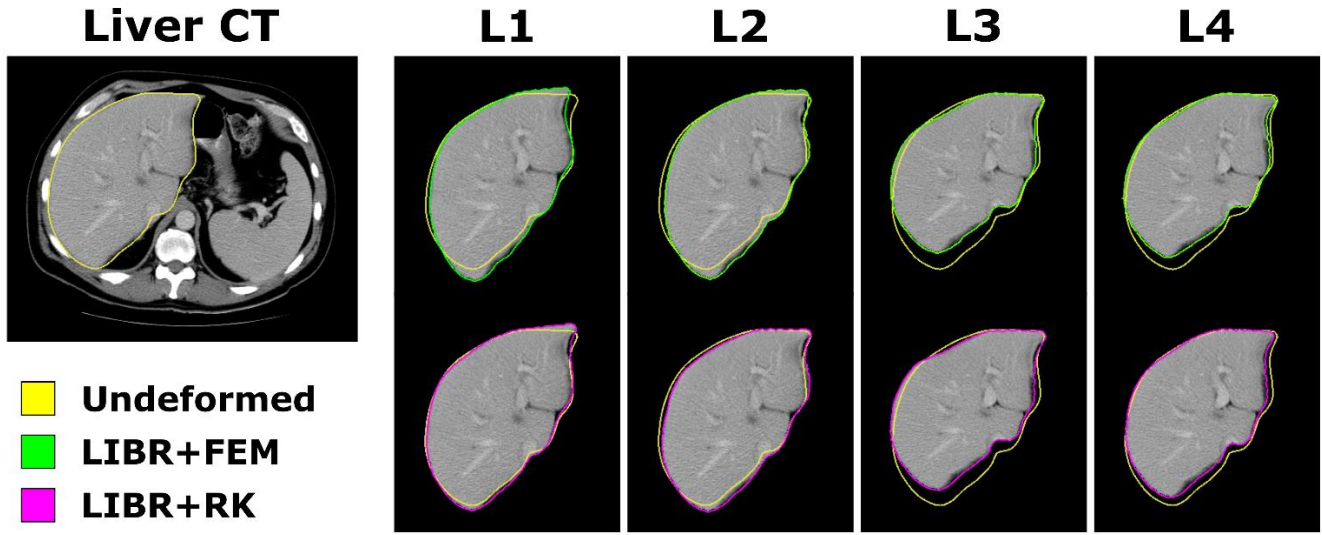

*Figure 8 – Image deformation results from the phantom liver dataset. An axial slice with the undeformed segmented liver (yellow) shows the original anatomy. The LIBR+FEM and LIBR+RK methods are used to deform the volume and segmentation contours (green and purple) for 4 example data configurations (L1-L4).*

The *in vivo* breast dataset provides a unique validation dataset to compare reconstructed deformed images to a validation image. Image slice comparisons after registration with the LIBR+FEM and LIBR+RK methods from three example cases (B1-B3) are shown in Figure 9. The MR images acquired in the arm-down and the arm-up configurations both show breast glandular feature patterns. The arm-down image volume is deformed using the displacement fields from the optimal LIBR+FEM and the LIBR+RK registrations. The deformed image slices from both methods show general glandular feature agreement with the arm-up validation image slices. However, there are also misaligned features present in both images which are most noticeable in the largest volume case (B3). Looking at specific glandular features, feature 1 in B1 appears similar in the arm-up, LIBR+FEM, and LIBR+RK images. In B2, feature 2 shows strong agreement across all three images, but feature 3 presents differently between the LIBR+FEM and LIBR+RK images. Feature 4 in B3 shows the most discrepancy – it varies in shape between the arm-up, LIBR+FEM, and LIBR+RK images.

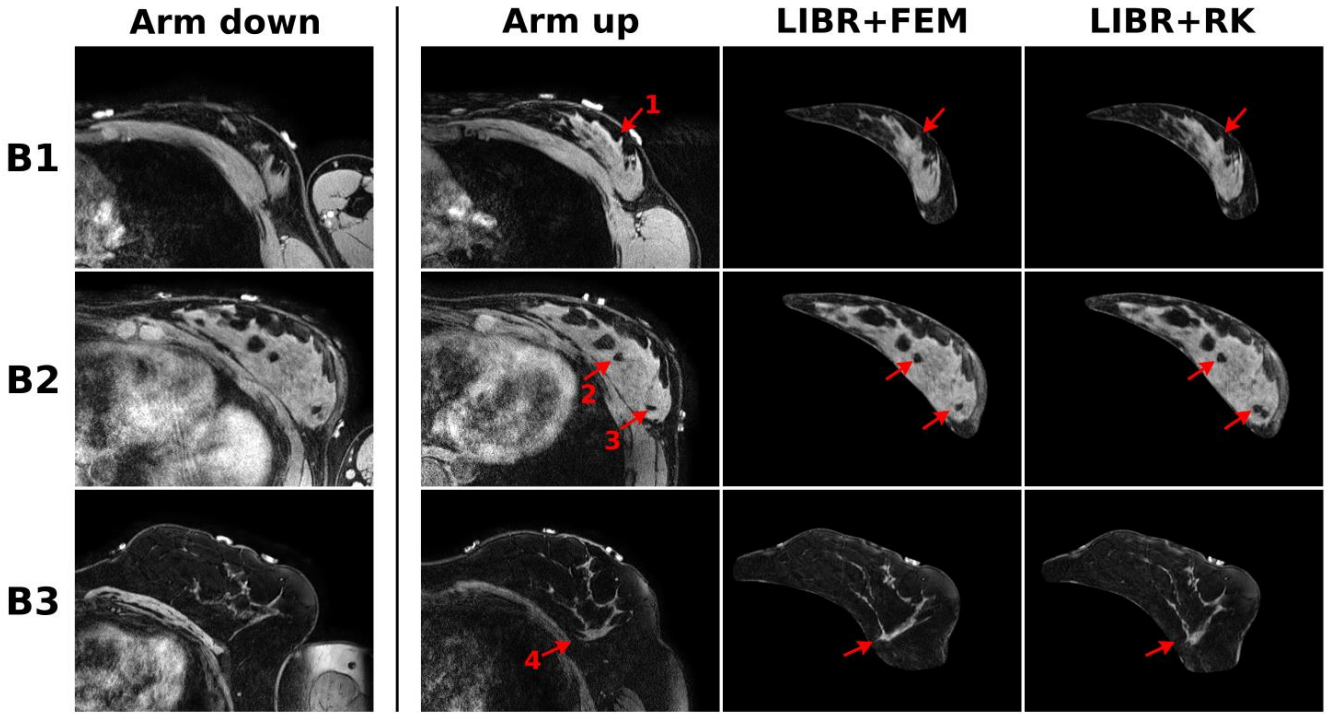

*Figure 9 – Image deformation results from the breast dataset of three example cases (B1-B3) shown as axial supine breast MR imaging slices. The acquired arm-down image column was used as the moving space and the acquired arm-up image column was used as the fixed space. Registered images using the LIBR+FEM method and the LIBR+RK method show general glandular feature agreement with the image slices in the arm-up column. Specific features (1-4) are denoted in red.*

## 4 Discussion

The results show that regularized Kelvinlets can be used for reconstructing deformation states on an elastic tissue-mimicking liver phantom dataset and on an *in vivo* breast imaging dataset. The accuracy of these registered deformation states is comparable to the accuracy of deformation states generated with a more traditional FEM method. Previous applications of regularized Kelvinlets have been confined to creating realistic *in silico* deformations for 2D or 3D animation. This work evaluates regularized Kelvinlets for modeling elastic deformations in the physical world. Regularized Kelvinlet deformations are computed in an infinite elastic domain that is not representative of deformations for physical elastic objects with finite material bounds. Although this is a limitation, the results here demonstrate that regularized Kelvinlets can offer computational advantages without significant degradation in accuracy specifically for sparse data organ registration problems. One implementation consideration is the selection of the regularization radial scale $\varepsilon$ and number of control points $k$ parameters for a registration application where validation data is not available. Registration accuracy demonstrated little degradation (<1 mm) within a range of $\varepsilon$ and $k$ values as shown in Figure 5, showing that accuracy performance is robust for varying parameter values. Because the $k$ parameter is directly related to precomputation and reconstruction computation time, $k=40$ is likely an adequate choice to balance between accuracy and computation time. For the regularization radial scale parameter $\varepsilon$, the range that provides approximately equivalent accuracy shown in Figure 5 (2 mm – 20 cm) is quite broad. A proposed $\varepsilon$ range of 2–5 cm is likely sufficient for other registration applications, assuming a similar geometric scale to liver and breast. However, the $\varepsilon$ parameter controls the smoothness of the superposed regularized point loads and the degree of subsurface field penetration. These factors should be considered on a case-by-case basis when applying this method to new applications.

By utilizing analytic solutions for linear elasticity, regularized Kelvinlets offer significant computational advantages compared to the FEM method. As shown in Figure 4, total computation times for registration using regularized Kelvinlets were reduced on average by amounts ranging from 86 seconds (breast dataset with 10 control points) to 34 minutes (liver dataset with 100 control points). While much of this computation time can be precomputed, reduction of precomputation burden directly improves practical utility of the LIBR method. Using regularized Kelvinlets also eliminates the need for large matrix assembly and inversions required for the FEM method. In this work, both the LIBR+FEM and LIBR+RK methods utilized tetrahedral meshes. However, the only purpose of the mesh for the LIBR+RK method was for computing the comparable strain energy regularization term used in the LIBR+FEM method. A mesh is not actually required for generating displacements using regularized Kelvinlets. An unstructured point cloud is sufficient since only the distance to the control point forcing location is needed to compute displacement per equation (4). Thus, the LIBR+RK method could be implemented as a meshless method in future work with an alternative regularization approach, further reducing the complexity associated with 3D mesh generation. Additionally, both methods could be multithreaded or implemented on a GPU for faster computation times. The speed and complexity advantages associated with regularized Kelvinlets could be beneficial for applications where real-time deformation modeling is desirable like for intraoperative image guidance or surgical training simulators.

The accuracy of both registration methods is generally comparable, with the average optimal RMS TRE values ranging between 3-7 mm for both datasets shown in Figure 5. For the liver dataset, the optimal RMS TRE values and the individual target errors for the 4 example cases (L1-L4) shown in Figure 7 were lower when using the LIBR+FEM method with 40 control points distributed on the organ's posterior surface than when using the LIBR+RK method with 160 control points and $\varepsilon = 0.01$ distributed on the entire organ's surface. The Sparse Data Challenge liver dataset was designed as a controlled experiment with well-defined materials and loading conditions. The phantom was composed of a linear elastic silicone material, and deformations were caused by adding and removing padding under the posterior surface of the liver phantom. The LIBR+FEM method models the unique geometry of the liver phantom. Placing control points on the posterior surface with the LIBR+FEM method designates the anterior surface of the organ as stress free and accurately represents the experimental conditions. The LIBR+RK method models deformations in an infinite elastic domain and cannot represent the experimental conditions as accurately as the LIBR+FEM method, which may be a cause of the degradation in TRE results.

Contrarily, for the breast dataset, the optimal RMS TRE and individual target errors for the 3 example cases (B1-B3) shown in Figure 7 were higher when using the LIBR+FEM method with 40 control points distributed on the organ's posterior surface than when using the LIBR+RK method with 70 control points and $\varepsilon = 0.05$ distributed on the organ's posterior surface. The breast dataset was generated from supine MR images of healthy volunteers' breasts, and the deformations were caused by the volunteers' arm motion. The composition of breast tissue is much more heterogeneous than that of a silicone phantom, given that it contains adipose tissue, glandular tissue, muscle, skin, ligaments, and fascia. Additionally, the loading conditions causing the deformations are more ambiguous and more difficult to model with FEM boundary conditions. In this work, it is interesting that the LIBR+RK method improves TRE performance compared to the LIBR+FEM method for the breast dataset despite assuming an infinite linear elastic domain. This may be due to the breast not having well defined inferior or superior boundaries meaning that the segmented breast geometry is less informative for modeling the global deformations. This is not meant to discount the idea that a more complex FEM model would likely be able to more accurately model deformations from arm motion and improve TRE results. Rather, it suggests that with limited sparse registration data and ambiguous loading conditions, the LIBR+RK method offers equivalent (or even slightly improved) registration accuracy compared to the LIBR+FEM method.

The comparability of both methods is further emphasized by comparing the image slices deformed with both methods in Figures 8 and 9. The images deformed with both methods would offer similar information when used in an image guided surgery application, with realistic underlying deformation bases that do not corrupt the original image quality. For the breast dataset, the glandular features of the breast tissue deformed with the LIBR+FEM method and the LIBR+RK method are both qualitatively similar to the image acquired in the arm-up position which is encouraging for the intraoperative registration application.

Finally, the regularized Kelvinlets method presented here made use of the "grab" regularized Kelvinlets solution to model deformations. Utilizing other regularized Kelvinlets derived deformations for locally affine loads described in (de Goes and James, 2017) – namely the "pinch", "scale", and "twist" deformations – may be applicable for modeling other types of deformations in medical data. Since the two datasets used here undergo large scale whole organ deformations, the "grab" regularized Kelvinlets solutions were selected as the most appropriate for modeling these deformations. However, for other datasets that feature more local deformations like the deformations caused by tissue-tool interactions, these other types of regularized Kelvinlets sculpting brushes may be applicable.

## 5 Conclusion

Regularized Kelvinlets are used as a biomechanical model for linear elasticity that offer equivalent performance with better computation time for sparse data registration compared to an equivalent method that uses the finite element method. This proposed algorithm using regularized Kelvinlets for registration was demonstrated on a phantom dataset and an *in vivo* deformation dataset for two different soft tissue organs, demonstrating its utility and generalizability to multiple organ registration problems. A regularized Kelvinlets model, with its ease of implementation and reduced encumbrance, offers distinct computational advantages that realize a fundamental step forward in enabling nonrigid, near real-time correction for image-guided surgery applications.

**Acknowledgements**

This work was supported by the National Institutes of Health through Grant Nos. R01EB027498 and T32EB021937, the National Science Foundation for a Graduate Research Fellowship awarded to M.J.R., and the Vanderbilt Center for Human Imaging supported by Grant No. 1S10OD021771-01 for the 3T MRI.